# High-speed linear optics quantum computing using active feed-forward

Robert Prevedel[1], Philip Walther[1,2], Felix Tiefenbacher[1,3], Pascal Böhi[1,*], Rainer Kaltenbaek[1], Thomas Jennewein[3] & Anton Zeilinger[1,3]

[1] *Institute for Experimental Physics, University of Vienna, Boltzmanngasse 5, A-1090 Vienna, Austria*

[2] *Physics Department, Harvard University, Cambridge, Massachusetts 02138, USA*

[3] *Institute for Quantum Optics and Quantum Information (IQOQI), Austrian Academy of Sciences, Boltzmanngasse 3, A-1090 Vienna, Austria*

\* *Present address: Max-Planck-Institut für Quantenoptik und Sektion Physik der Ludwig-Maximilians-Universität, Schellingstr. 4, 80799 München, Germany*

**As information carriers in quantum computing[1], photonic qubits have the advantage of undergoing negligible decoherence. However, the absence of any significant photon–photon interaction is problematic for the realization of non-trivial two-qubit gates. One solution is to introduce an effective nonlinearity by measurements resulting in probabilistic gate operations[2,3]. In one-way quantum computation[4–8], the random quantum measurement error can be overcome by applying a feed-forward technique, such that the future measurement basis depends on earlier measurement results. This technique is crucial for achieving deterministic quantum computation once a cluster state (the highly entangled multiparticle state on which one-way quantum computation is based) is prepared. Here we realize a concatenated scheme of measurement and active feed-forward in a one-way quantum computing experiment. We demonstrate that, for a perfect cluster state and no photon loss, our quantum computation scheme would operate with good fidelity and that our feed-forward components function with very high speed and low error for detected photons. With present technology, the individual computational step (in our case the individual feed-forward cycle) can be operated in less than 150 ns using electro-optical modulators. This is an important result for the**

**future development of one-way quantum computers, whose large-scale implementation will depend on advances in the production and detection of the required highly entangled cluster states.**

One-way quantum computation is based on highly entangled multi-particle states, so-called cluster states, which are a resource for universal quantum computing. On these cluster states, single-qubit measurements alone are sufficient to implement universal quantum computation. Different algorithms only require a different "pattern" of single-qubit operations on a sufficiently large cluster state; as explained in ref. 5, "the cluster states are one-way quantum computers and the measurements form the program." In contrast to the standard linear optics architecture, which relies on multi-particle gates, the cluster state computation is performed by consecutive single-qubit measurements where the choice of the future measurement basis is dependent on the outcome of preceding measurements. Active feed-forward of the classical measurement results renders one-way quantum computation *deterministic,* i.e. given a perfect cluster state and exact measurements, the processing of encoded information on physical qubits is accomplished without error. The one-way quantum computer model that we employ is currently the only one which promises deterministic photonic quantum computation (through feed-forward). Standard optical schemes[2] achieve this only in the asymptotic regime of numerous gates and/or photons. Nevertheless, we note that feed-forward control based on measurements made on ancillary qubits is also essential for error correction in the standard network approach.

Recently, the working principles of one-way quantum computing have been demonstrated using cluster states encoded into the polarization states of photons[9-11]. However, all experiments so far have been performed using fixed polarizer settings, thus making the computation probabilistic (i.e. not scalable) and wasting precious resources on the way. In this Letter we demonstrate feed-forward linear-optics quantum computation on a four-qubit cluster state. The cluster state creation is based on a post-selection technique developed in Ref.(9) and the feed-forward stages are realized by employing fibre delays and fast active switches for selecting the appropriate measurement basis and correcting introduced Pauli errors. Earlier proof-of-principle demonstrations[12-14] of feed-forward control were limited to two photons and one feed-forward step only. However, to demonstrate feed-forward quantum computing, more photons and thus several consecutive feed-forward steps are required. It is particularly important to realize a situation where a later measurement depends on an earlier measurement and its fed forward result. Dealing with the complex situation of a four-qubit cluster state and three EOMs, we demonstrate "error-free" single-qubit and two-qubit gate operations as well as Grover's search algorithm[15].

Given a cluster state, two basic types of single-particle measurements suffice to operate the one-way quantum computer. Measurements in the computational basis $\{|0\rangle_j, |1\rangle_j\}$ have the effect of disentangling, i.e., removing the physical qubit $j$ from the cluster, thus leaving a smaller cluster state. The measurements which perform the actual quantum information processing, however, are made in the basis $B_j(\alpha) = \{|\alpha_+\rangle_j, |\alpha_-\rangle_j\}$, where $|\alpha_\pm\rangle_j = \left(e^{i\alpha/2}|0\rangle_j \pm e^{-i\alpha/2}|1\rangle_j\right)/\sqrt{2}$ with $\alpha \in \{0, 2\pi\}$. The choice of measurement basis determines the single-qubit rotation, $R_z(\alpha) = \exp(-i\alpha\sigma_z/2)$, followed by a Hadamard operation, $H = (\sigma_x + \sigma_z)/\sqrt{2}$, of encoded qubits in the cluster[9] ($\sigma_x, \sigma_y, \sigma_z$ being the Pauli matrices). Any quantum logic operation can be carried out by an appropriate choice of $B_j(\alpha)$ on a sufficiently large cluster state. We define the outcome $s_j$ of a measurement on the physical qubit $j$ to be "0" if the measurement outcome is $|\alpha_+\rangle_j$ and "1" if the outcome is $|\alpha_-\rangle_j$. Whenever the outcome is "0", the computation proceeds without error, while for the case where the outcome is "1", a well-defined Pauli error is introduced. These known errors are compensated for by feed-forward such that the output controls future measurement settings.

In the present work, we create a cluster state of the form,

$$|\Phi_{cluster}\rangle = \frac{1}{2}\left(|0\rangle_1|0\rangle_2|0\rangle_3|0\rangle_4 + |0\rangle_1|0\rangle_2|1\rangle_3|1\rangle_4 + |1\rangle_1|1\rangle_2|0\rangle_3|0\rangle_4 - |1\rangle_1|1\rangle_2|1\rangle_3|1\rangle_4\right) \quad (1),$$

where $|0\rangle$ and $|1\rangle$, in the actual experiment, denote horizontal and vertical polarization, respectively (the subscript labels the photon). The state of Eq. (1) is equivalent to the four-qubit linear cluster $|\Phi_{\text{lin}4}\rangle$ and to the horseshoe cluster[9] $|\Phi_{\subset 4}\rangle$ under the local unitary operation $H_1 \otimes I_2 \otimes I_3 \otimes H_4$ on the physical cluster state ($H$ is the Hadamard and $I$ is the identity operation). In the experiment, the cluster state is known to have been prepared when all four photons are detected. This ensures that photon loss and photo-detector inefficiency do not affect the experimental results. The state creation is verified by over-complete state tomography in which the density matrix of the cluster state's is reconstructed from a set of 1296 local measurements using a maximum-likelihood technique[16,17] and all combinations of mutually unbiased basis sets for individual qubits, i.e. $\{|0\rangle,|1\rangle;|+\rangle,|-\rangle;|R\rangle,|L\rangle\}$, where $|+/-\rangle = (|0\rangle \pm |1\rangle)/\sqrt{2}$ denote ±45° polarization and $|L/R\rangle = (|0\rangle \pm i|1\rangle)/\sqrt{2}$ stands for right and left circular polarization. Each of these measurements took 500 seconds. The experimentally obtained density matrix, $\rho$, has a fidelity of $F = \langle \Phi_{cluster} | \rho | \Phi_{cluster} \rangle = (0.62 \pm 0.01)$ with the ideal four-qubit cluster, $|\Phi_{cluster}\rangle$, which despite all EOMs and fibre-coupled outputs is sufficiently above the threshold for entanglement[18] of 0.5.

Using present technologies and customized fast EOMs, we were able to realize high-fidelity (>99% for detected photons) fast active switching with feed-forward times of less than 150 ns (cf. Methods). The therefore achievable gate-operation is about three orders of magnitude faster than comparable physical realizations of quantum computers[19-21]. The measurement device for an arbitrary basis consists of a quarter-wave and a half-wave plate followed by a polarizing beam-splitter (PBS), which transmits horizontally polarized light ("0") and reflects vertically polarized light ("1"). While qubit 1 and qubit 2 are measured without any delay, qubit 3 and qubit 4 are delayed in optical single-mode fibres with lengths of 30m (150 ns) and 60m (300 ns), respectively. The active switching itself is achieved via Pockels cells; one for qubit 3 to adapt the measurement basis, i.e. from $B_3(\beta)$ to $B_3(-\beta)$, and two in the channel of output-qubit 4 to correct introduced Pauli-errors, $\sigma_x$ and $\sigma_z$ (see Fig. 1).

As an example, consider the general case of a three-qubit linear cluster state $|\Phi_{lin3}\rangle$, such as the one depicted in Figure 2(a). This state can be obtained from our four-qubit cluster by removing qubit 1, i.e. measuring this qubit in the computational basis for the linear cluster, $\{|+\rangle_1,|-\rangle_1\}$. Consecutive measurements in bases $B_2(\alpha)$ and $B_3(\beta)$ on the physical qubits 2 and 3 implement an arbitrary single-qubit rotation of the encoded input qubit $|\Psi_{in}\rangle=|+\rangle_1$. These measurements rotate the encoded input qubit to the output state $|\Psi_{out}\rangle=\sigma_x^{s_3}HR_z((-1)^{s_2}\beta)\sigma_x^{s_2}HR_z(\alpha)|\Psi_{in}\rangle=\sigma_x^{s_3}\sigma_z^{s_2}R_x((-1)^{s_2}\beta)R_z(\alpha)|\Psi_{in}\rangle$, which is stored on qubit 4. The measurement outcome, $s_i=\{0,1\}$, on the physical qubit $i$, (1) determines the measurement basis for the succeeding qubit and (2) indicates any introduced Pauli errors that have to be compensated for. In the specific case where the outcomes of the second and third qubit are $s_2=s_3=0$, no error correction is required: $|\Psi_{out}\rangle=R_x(\beta)R_z(\alpha)|\Psi_{in}\rangle$. Whenever the outcome of the second qubit is $s_2=1$ ($s_3=0$), then the measurement basis of the third qubit has to be changed from $B_3(\beta)$ to $B_3(-\beta)$ and finalized by a Pauli error correction, $\sigma_z$, i.e. $|\Psi_{out}\rangle=\sigma_z R_x(-\beta)R_z(\alpha)|\Psi_{in}\rangle$ to get the same output as if no error had occurred. Similar corrections are required in the cases when the third qubit's outcome is $s_3=1$ ($s_2=0$): $|\Psi_{out}\rangle=\sigma_x R_x(\beta)R_z(\alpha)|\Psi_{in}\rangle$ or, if an unwanted projection occurs to both qubits, $s_2=s_3=1$, two Pauli error, $\sigma_z$ and $\sigma_x$, have to be compensated for on qubit 4, $|\Psi_{out}\rangle=\sigma_x\sigma_z R_x(-\beta)R_z(\alpha)|\Psi_{in}\rangle$. The same feed-forward techniques hold for two-dimensional cluster-states (Figure 3) which will be discussed later in the paper.

Examples of single-qubit rotations with feed-forward are shown in Figure 2, together with the outcomes of the same computation in the case when no feed-forward is applied. In each case the output of the single-qubit rotation is stored in qubit 4 and completely characterized by single-qubit tomography. Figure 2(a) shows a schematic of the implemented quantum algorithm; Figure 2(b) shows the output of the computation $|\Psi_{out}\rangle=R_x\left(-\frac{\pi}{2}\right)R_z\left(-\frac{\pi}{2}\right)|\Psi_{in}\rangle=|+\rangle_4$, in the laboratory basis, with and without active feed-forward. We find an average fidelity of $(0.84\pm0.08)$ with the ideal state when active feed-forward is implemented. This is a considerable improvement over the case of no feed-forward, which produces the target state with an average fidelity of only $(0.55\pm0.06)$. In order to prove universal quantum computing we need to demonstrate single-qubit rotations outside the Clifford group[22]. This special example is shown in Figure 2(c), where we perform polarization projections in the basis $\alpha=\frac{\pi}{4}$ and $\beta=\frac{\pi}{12}$ which results in the more complex computation $|\Psi_{out}\rangle_4 = R_x\left(\frac{\pi}{4}\right)R_z\left(\frac{\pi}{12}\right)|\Psi_{in}\rangle = \cos\left(\frac{\pi}{8}\right)|H\rangle + \sin\left(\frac{\pi}{8}\right)e^{i\pi/12}|V\rangle$ in the error-free case. Here we find an average fidelity of $(0.79 \pm 0.07)$ with active feed-forward, but only $(0.45 \pm 0.05)$ without (see Fig. 2). We find similar results for other measurement angles and hence other single-qubit rotations. The error bars of the above results were estimated by performing a 100 run Monte Carlo simulation of the whole state tomography analysis, with Poissonian noise added to the count statistics in each run. Note that the reduced output state fidelity is mainly due to the non-ideal cluster state preparation and not due to erroneous switching of the EOMs, as these operate with very high precision.

It is a specific strength of the cluster-state computation that the adaptation of the measurement basis, $B_j(\alpha)$, caused by the measurement outcome of the preceding qubit, can be carried out without active switching when the eigenstates of the measurement basis are identical to the eigenstates of $\sigma_x, \sigma_y$ or $\sigma_z$. In that case logical feed-forward results in a reinterpretation of the measurement outcome. Outcome "0" would then correspond to the measurement outcome $|\alpha_-\rangle_j$ and "1" to the outcome $|\alpha_+\rangle_j$. We demonstrate this specific feature within the 2-dimensional four-qubit cluster-states, the horse-shoe cluster, $|\Phi_{\subset 4}\rangle$, and the box cluster, $|\Phi_{\square 4}\rangle$, which we use to realize an entangling gate and an implementation of Grover's quantum search algorithm.

Universal quantum computing requires a universal set of one- and two-qubit operations such as the controlled-NOT (CNOT) or controlled-PHASE (CPhase) gates which can be realized using either horseshoe- or box cluster. These gates can be implemented on our linear cluster by changing the order of measurements, e.g. by measuring qubits 2 and 3 and thus transferring the two-qubit quantum state onto the remaining qubits 1 and 4. This quantum circuit can also be written as $|\Psi_{out}\rangle = \left(\sigma_x^{s_2} \otimes \sigma_x^{s_3}\right)(H_1 \otimes H_2)[R_z(\alpha) \otimes R_z(\beta)]CPhase|\Psi_{in}\rangle$ where $|\Psi_{in}\rangle = |+\rangle_1|+\rangle_2$ is our encoded two-qubit input state. Note that the Pauli errors have to be compensated for in the case where $s_2 = 1$ and/or $s_3 = 1$. In principle, feed-forward relations in the case of two-qubit gates are more complex, as measurement errors in one "rail" may influence the state of the qubit in another rail[23]. In particular, for polarization projections

$|\alpha_+\rangle_2|\beta_+\rangle_3, |\alpha_+\rangle_2|\beta_-\rangle_3, |\alpha_-\rangle_2|\beta_+\rangle_3, |\alpha_-\rangle_2|\beta_-\rangle_3$, (i.e. for measurement outcomes $s_2 = s_3 = 0; s_2 = 0, s_3 = 1; s_2 = 1, s_3 = 0; s_2 = s_3 = 1$), the operation $I_1 \otimes I_4, I_1 \otimes \sigma_{x4}, \sigma_{x1} \otimes I_4, \sigma_{x1} \otimes \sigma_{x4}$ has to be fed-forward to the output qubits 1 and 4, respectively. In Figure 3, we explicitly show the case where both photons 2 and 3 are measured to be $s_2 = s_3 = 1$ instead of the desired "0" outcomes $s_2 = s_3 = 0$ in the bases $B_2(0)$ and $B_3(0)$. Those "errors" rotate the input state to the maximally entangled output state $|\Psi_{out}\rangle_{1,4} = \frac{1}{\sqrt{2}}(|+\rangle_1|V\rangle_4 - |-\rangle_1|H\rangle_4)$. To obtain the desired state $|\Psi_{out}\rangle_{1,4} = \frac{1}{\sqrt{2}}(|H\rangle_1|+\rangle_4 + |V\rangle_1|-\rangle_4)$, the operation $\sigma_{x1} \otimes \sigma_{x4}$ has to be fed-forward on qubits 1 and 4. Density matrices of the ideal two-qubit output state and the experimentally reconstructed state are shown in Figure 3, together with the measured output state obtained without feed-forward. We compute a state fidelity of $(0.79 \pm 0.04)$ for the overlap of our experimental fed-forward state with the ideal one. The tangle[24] of this output state is $\tau = (0.42 \pm 0.09)$, confirming the generation of entanglement between the output qubits as a result of the computation. Furthermore, our reconstructed density matrix implies a maximum CHSH Bell parameter[25] of $S = (2.40 \pm 0.09)$, which is more than four standard deviations above the $S = 2$ upper limit for local realistic theories. For comparison, if the feed-forward relation is not applied to this specific computation, the measured fidelity is only $(0.09 \pm 0.03)$, in agreement with the theoretical prediction of 0 – no overlap with the desired state.

Quantum algorithms[15,26] are fascinating applications of quantum computers. Interestingly, Grover's quantum search algorithm[9,15,27] can be implemented on a four-qubit box cluster, such as the one depicted in Figure 4(a), with final readout measurements made in the basis $B_{1,4}(\pi)$ on physical qubits 1 and 4. Grover's algorithm promises a quadratic speedup for unstructured search. It is worth mentioning that, for the case of four entries in an unsorted database[15], Grover' search will find the marked entry with certainty after a single iteration. The algorithm can be separated into two basic operations. First, a quantum device – the so-called 'oracle' – labels the correct element, which can be set by a proper choice of $\alpha$ and $\beta$, specifically, it tags one of the four computational basis states $|0\rangle|0\rangle$, $|0\rangle|1\rangle$, $|1\rangle|0\rangle$ and $|1\rangle|1\rangle$ by changing its sign, e.g. $|0\rangle|0\rangle \rightarrow -|0\rangle|0\rangle$. Then after an inversion-about-the-mean operation[9,15] the labelled element is found with certainty by the readout measurements. However, incorrect measurement outcomes at the 'oracle' (i.e. at qubits 2 and 3) introduce Pauli-errors, which effectively cause a wrong database element to be tagged. Feed-forward compensates for these errors such that the algorithm produces the right search result with certainty. In Fig. 4(b), we show the experimental results of this quantum algorithm with and without feed-forward. The difference in performance is quite obvious, with feed-forward the right database element is identified with a probability of $(85 \pm 3)\%$, which compares favourably with the case when the feed-forward relation is not applied, which we find to be $(25 \pm 2)\%$, just as good as with a classical random search algorithm.

In summary, we have shown that in the absence of photon loss, a one-way quantum computer with active, concatenated feed-forward would operate with high fidelity. As in all current photonic quantum computation experiments, the input cluster state is produced conditional on detecting all constituent photons. Because the efficiency of producing cluster states is low at present, this is not yet scalable. However, the technique is insensitive to photon loss due to absorption, reflection, fibre coupling and photo-detector inefficiency. Thus our experiments show that except for photon loss, the feed-forward procedure operates with a quality and speed presently unmatched by other quantum computation methods. Conceptually the most interesting result of our work is that it is indeed possible to build a deterministic quantum computer that has intrinsically random quantum measurements as essential feature. Eventual large-scale implementations will need significant improvements of state preparation quality and photon detection efficiency, and reduction of photon losses. This will certainly be fostered by recent developments of highly efficient single-photon detection methods as well as "on demand" single-photon sources. Given large and high-fidelity cluster states as well as low photon loss and significantly improved detectors, promising future applications of one-way quantum computers include important tasks such as the quantum Fourier transform[28,29] which is at the heart of Shor's factorizing algorithm.

**Methods**

**Experimental cluster state preparation**

In our experiment, the cluster state is generated from photon pairs entangled in polarization and mode which originate from spontaneous parametric down-conversion. We employed the method of Ref. (9) to generate the four-qubit cluster state which is shown in Fig. 1 together with its extension to realize feed-forward quantum computation. The generation of the cluster state is dependent on simultaneous emission of four photons, i.e. we only post-select those cases where each the four output modes a-d of the polarizing beam-splitters contain one photon (for more details, see the Methods section of Ref. (9)).

**Contributions to feed-forward time**

Quantum computation on a cluster state is performed by consecutive measurements on qubits 1-4. It is therefore necessary to locally delay photons 3 and 4 if active feed-forward of measurement results is desired. We find that the overall process of a single feed-forward step requires, on average, 145±3 ns, where this value is composed of the following contributions: propagation time of photons 1 and 2 in single-mode fibres leading to detector (15 ns), delay of the single-photon detectors (35±3 ns), processing time of the logic (7.5 ns), switching delay of the EOM driver (65 ns), rise time of the Pockels cell (5 ns), and miscellaneous coaxial cables employed in the set-up (17.5 ns). Two single-mode fibres, 30 m (150 ns) and 60 m (300 ns) long, serve to locally delay the photons during the detection stage, logic operations and the switching/charging process of the EOMs. We expect that the overall feed-forward time can significantly be reduced in optical fibres or waveguides where a smaller scale results in faster switching of the EOM driver[30].

**Characterisation of the feed-forward stage**

For the active switching, we employ KD*P (potassium dideuterium phosphate) crystals with a measured transmission greater than 96%, a half-wave voltage of ~6.3 kV and a high switching contrast of approximately 500:1. The switching contrast is defined as the ratio of photons that are measured to obtain a well-defined polarization rotation due to the operation of the EOM divided by the photons that remain in the original state due to malfunctioning of the device. This was measured on the single photon level for various input polarizations employing time-correlated photons emitted by a down-conversion source, triggering on one photon and thereby rotating the polarization state of the other photon. From the high switching contrast of 500:1, one can infer that the total feed-forward accuracy of the three EOMs for detected photons is at least $(1-1/500)^3 > 0.99$. Other errors apart from photon loss such as mode-mismatch and unwanted phase-shifts at the main PBSs only result in non-ideally prepared input cluster states. However, the performance of the feed-forward stage is unaffected by these imperfections. In the present configuration, the custom-built EOM drivers can be operated with up to 20 kHz, compatible with our trigger-rate requirement which is set by our photon pair production rates (~2 kHz). During recharge cycles, the EOM-drivers are "disabled" for an effective dead time of 1.6 μs, which is short enough considering our average two- and four-photon production rate (on the order of 2 kHz and 1 Hz, respectively). The overall detection efficiency of the experiment bearing in mind the non-ideal collection of photons in single-mode fibres (~20%), quantum efficiency of the detectors (~55%) and various losses in fibres, optical elements and EOMs (~5%) is roughly 10% per detector, a standard figure in many multi-photon down-conversion experiments.

**Acknowledgements** We are grateful to M. Aspelmeyer, J.I. Cirac, J. Kofler, and K. Resch for discussions as well as to T. Bergmann and G. Mondl for assistance with the electronics. R.P. thanks E.-M. Röttger for assistance in the laboratory. We acknowledge financial support from the Austrian Science Fund (FWF), the European Commission under the Integrated Project Qubit Applications (QAP) funded by the IST directorate and the DTO-funded U.S. Army Research Office.

**Competing interests statement** The authors declare that they have no competing financial interests.

**Correspondence** and requests for materials should be addressed to R.P. (robert.prevedel@univie.ac.at) or A.Z. (zeilinger-office@quantum.at).

**Figure 1 Schematic drawing of the experimental set-up.** Interferometric cluster-state preparation is shown in **a**, and its extension to achieve active feed-forward of the one-way quantum computation is depicted in **b**. An ultraviolet (UV) laser pulse passes twice through a nonlinear crystal to produce polarization-entangled photon pairs in both the forward and backward direction. Compensators (Comp.) are half-wave plates (HWP) and BBO crystals used to counter walk-off effects in the down-conversion crystal. They are aligned such that $|\Phi^-\rangle$ and $|\Phi^+\rangle$ states are emitted in the forward and backward direction, respectively. Taking into account the possibility of double-pair emission and the action of the polarizing beam splitters (PBSs), the four amplitudes of the linear cluster state can be generated with an additional HWP in mode a. Once this is achieved, the computation proceeds by consecutive polarization measurements on photons 1–4. Dependent on the outcomes of photons 1–3, three fast electro-optical modulators (EOMs) are employed to implement the active feed-forward. One EOM adapts the measurement basis of photon 3, while two EOMs, aligned for $\sigma_x$ and $\sigma_z$ operation, apply the error correction on output photon 4. Two single-mode fibres, 30 m and 60 m long, serve to locally delay the photons during the detection stage, logics operation and switching/charging process of the EOMs. The polarization of the photons is measured by a PBS preceded by a HWP and a quarter-wave plate (QWP) in each mode.

**Figure 2 Active feed-forward of two different single-qubit rotations. a**, The linear three-qubit cluster state (obtained from our four-qubit cluster state) and the quantum circuit it implements. The operation $R_x(\alpha) = \exp(-i\alpha\sigma_x/2)$ can be implemented through the matrix identity $R_x(\alpha) = HR_z(\alpha)H$. **b, c**, Fidelity of the output state with the desired state in the case of active feed-forward and without feed-forward of measurement results. Both the experimentally measured fidelities (red bars) and the expected, ideal fidelities (grey bars) are given. It is immediately clear that, with feed-forward, the computation theoretically always produces the desired outcome with certainty, even if measurement outcomes in the $|\alpha\rangle_2, |\beta\rangle_3$ basis deviate from the desired $s_2 = s_3 = 0$ event ('++'). In **b**, $\alpha$ and $\beta$ were both set to $-\frac{\pi}{2}$, resulting in the output state $|\Psi_{out}\rangle_4 = |+\rangle$, while in **c**, the measurement angles were set to $\alpha = \frac{\pi}{4}$ and $\beta = \frac{\pi}{12}$. Averaged over all possible measurement outcomes, the overlap of the measured one-qubit density matrix with the ideal state with feed-forward is $(0.84 \pm 0.08)$ in **b** and $(0.79 \pm 0.07)$ in **c**, respectively. Without feed-forward, theory predicts an average fidelity of 0.5. In the experiment, we find $(0.55 \pm 0.06)$ and $(0.45 \pm 0.05)$, for **b** and **c**, respectively. Error bars indicate s.d.

**Figure 3 Feed-forward of a two-qubit operation.** We perform the operation $|+\rangle_{1E}|+\rangle_{2E} \rightarrow \frac{1}{\sqrt{2}}(|H\rangle_1|+\rangle_4 + |V\rangle_1|-\rangle_4)$ with single-qubit measurements in $B_2(0)$ and $B_3(0)$ carried out on photons 2 and 3 on the horseshoe cluster state $|\Phi_{\subset 4}\rangle$. **a**, The algorithm implemented by the horseshoe cluster. **b**, The ideal, expected density matrix, with the real part of the density matrix shown as a bar chart (upper figure) and

as a coloured grid plot (lower figure). The imaginary components of the density matrices are zero in theory and negligible in the experiment. **c**, In the case where photon 2 and 3's outcome was $s_2 = s_3 = 1$ instead of the desired '00' event, the logical feed-forward relation has been applied by relabelling the analyser output ports. Fidelity and measures of entanglement of the reconstructed the state can be inferred from the main text. In **d**, we show the output of the same quantum computation when no feed-forward is applied. The experimental density matrix in this case differs remarkably from the ideal one, which is reflected in the low state fidelity (see main text).

**Figure 4 Demonstration of Grover's search algorithm with feed-forward.** Here we chose to tag the $|0\rangle|0\rangle$ entry. **a**, The algorithm consists of two distinct operations: The 'oracle' tags the unsorted database element by measuring physical qubits 2 and 3 in the bases $B_{2,3}(\pi)$, while the inversion process finds the desired database entry with certainty after a single query. Owing to intrinsic measurement randomness, however, it happens with equal probability that other database entries become tagged. Without feed-forward, on average, this results in a balanced output of the algorithm, as can be seen from the experimental data in **b**. Applying the feed-forward procedure leads to an unambiguous search result, so that, on average, the algorithm finds the correct outcome with a probability of $(85 \pm 3)\%$. In the case without feed-forward, we find each possible result with equal probability of $(25 \pm 2)\%$. Error bars indicate s.d..

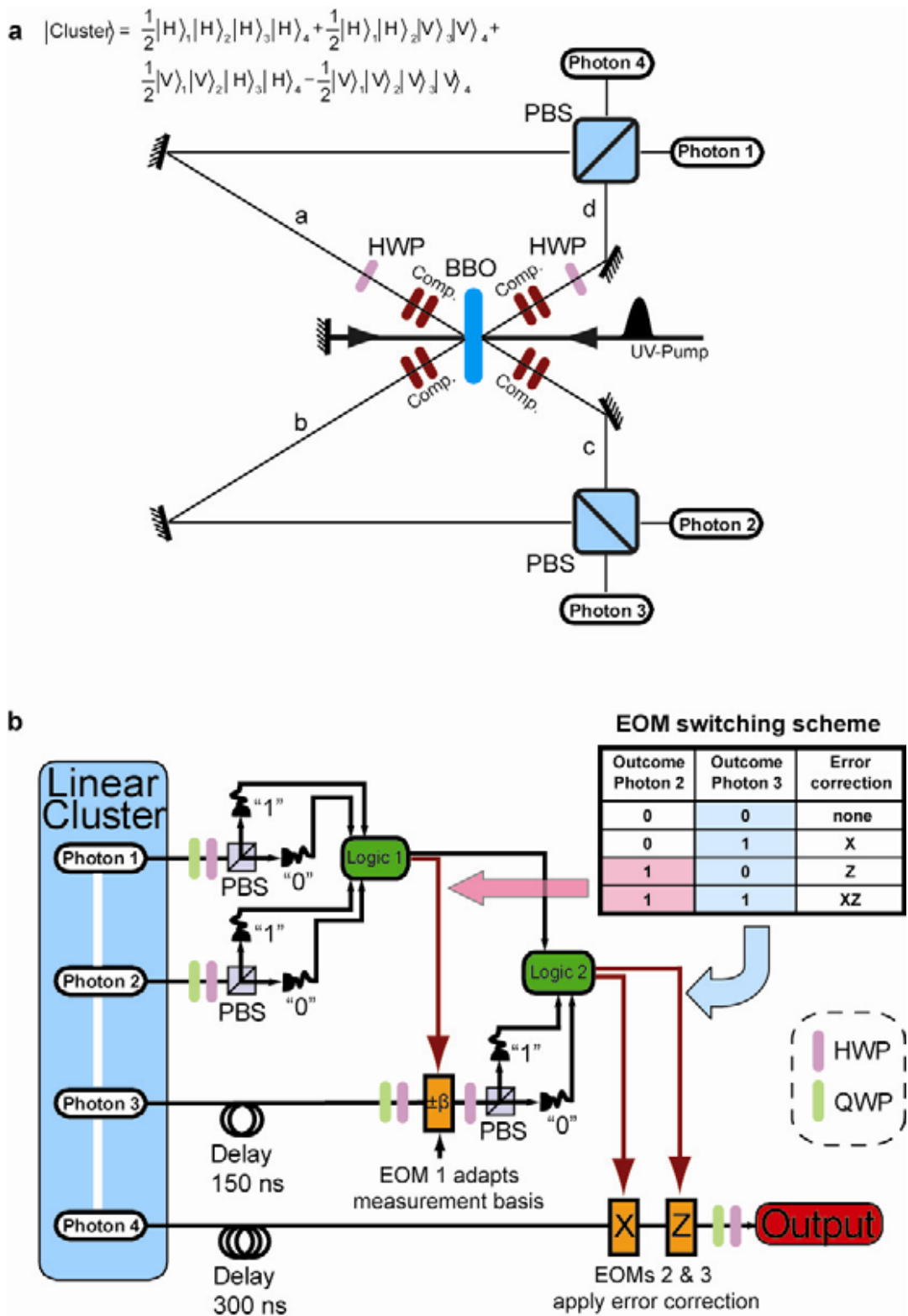

**Figure 1**

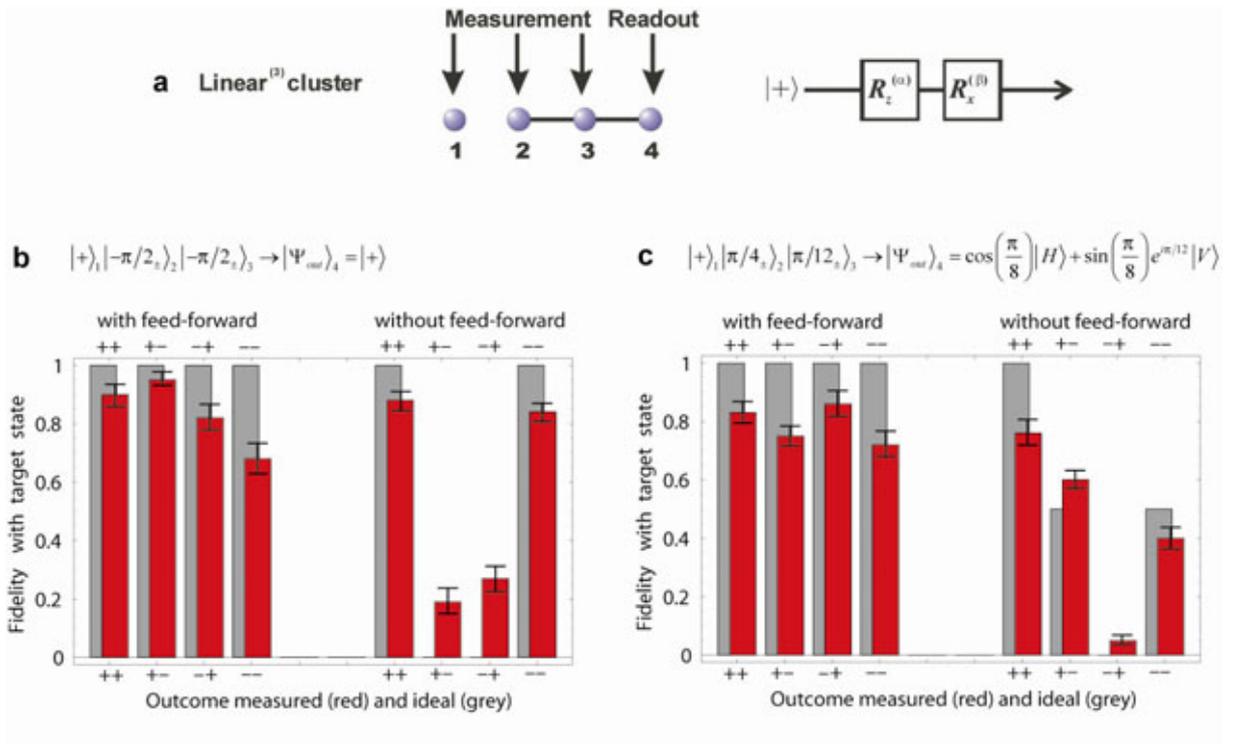

**Figure 2**

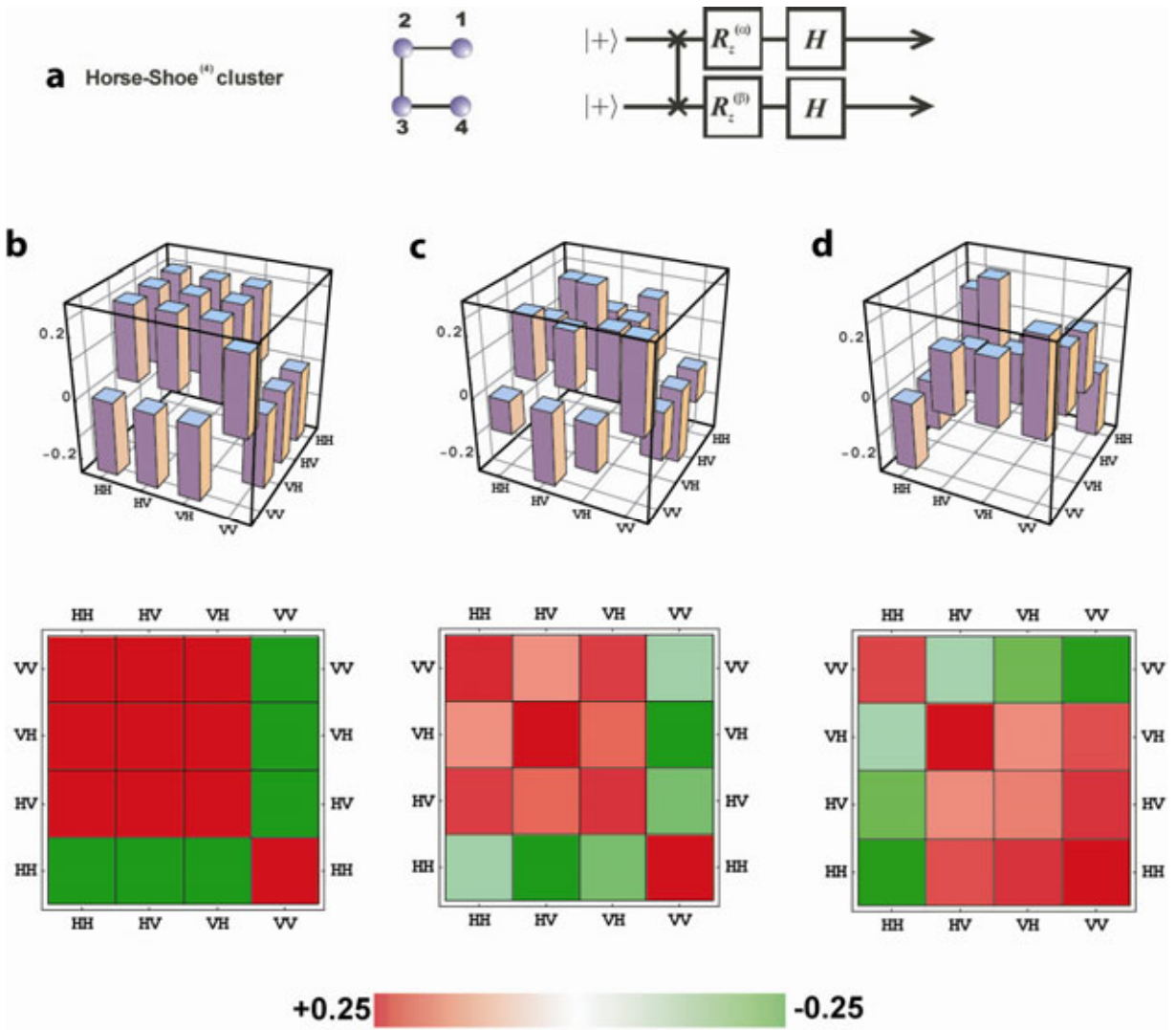

Figure 3

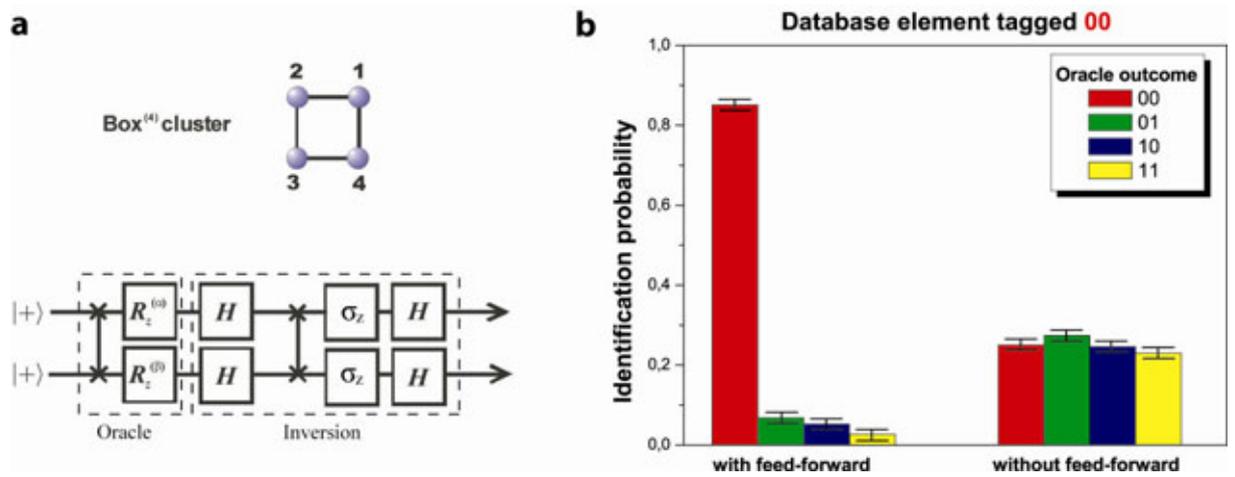

**Figure 4**